# OPTIMIZATION OF A WELDING PROCEDURE FOR MAKING CRITICAL ALUMINUM WELDS ON THE LBNF ABSORBER CORE BLOCK*

K. E. Anderson, A. Deshpande†, V. I. Sidorov, J. Zahurones
Fermi National Accelerator Laboratory, Batavia, IL, USA

*Abstract*

The LBNF Absorber consists of thirteen 6061-T6 aluminum core blocks. The core blocks are water cooled with deionized (DI) water which becomes radioactive during beam operations. The cooling water flows through gun-drilled channels in the core blocks. A weld quality optimization was performed to produce National Aeronautical Standard 1514 Class I quality welds on the aluminum core blocks. This was not successful in all cases. An existing Gas Tungsten Arc Welding Procedure Specification was fine tuned to minimize, in most cases, and eliminate detectable tungsten inclusions in the welds. All the weld coupons, however passed welding inspection as per the piping code: ASME B31.3 Normal Fluid Service. Tungsten electrode diameter, type, and manufacturer were varied. Some of the samples were pre-heated and others were not. It was observed that larger diameter electrodes, 5/32 in., with pre-heated joints resulted in welds with the least number of tungsten inclusions. It is hypothesized that thinner electrodes breakdown easily and get lodged into the weld pool during the welding process. This breakdown is further enhanced by the large temperature differential between the un-preheated sample and the hot electrode.

## OVERVIEW

LBNF Hadron Absorber is located downstream of the decay pipe. It consists of actively cooled aluminum and steel blocks surrounded by steel and concrete shielding [3, 4]. Cooling water is delivered to the core blocks via 2 in. 6061-T6 aluminum pipes that are directly welded to the blocks. These welds are inaccessible and are present in a high-radiation area. Thus, the welds must be of highest quality and adhere to suitable engineering code and last the duration of the experiment's lifetime. An existing welding procedure for small diameter aluminum pipe was fine-tuned to produce code quality welds.

## WELDING PROCEDURE

The welding procedure was applied to two types of critical welds that are present on the aluminum core blocks: pipe-block and plug welds. The base procedure involves performing meticulous cleaning and etching and handling operations on the 4043 filler rods. The rods are stored in a vacuum tube until the welding commences. All the parts to be welded are cleaned with isopropyl alcohol and are scraped to remove the oxide layer with a carbide scraping tool. An 80% Alternating Current Electro Negative (ACEN) setting is selected on the welding machine. A 99.9995% 75% helium and 25% argon gas mixture is used as the shielding gas. A 2% thoriated tungsten electrode between 3/32 in. and 5/32 in. (inclusive) is chosen. The purge gas flow is set at 130 SCFH (He.). The welding is done with one root pass followed by at least 2 additional passes. It must be noted that there are several additional fine-tuning steps in the procedure that are not highlighted here.

### Types of Weld Coupons

A joint welded as per a WPS and processed to be sent out for testing is referred to as a coupon. Multiple instances of two types of coupons were created: pipe-block and plug weld. A special assembly fixture was made to weld the plug weld coupons. The material of the fixture was 6061-T6 as well. The joint designs are shown in Figs. 1 and 2 below.

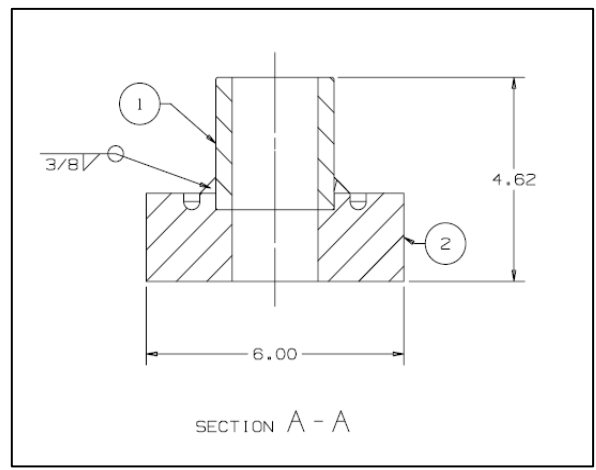

Figure 1: Pipe-block weld design.

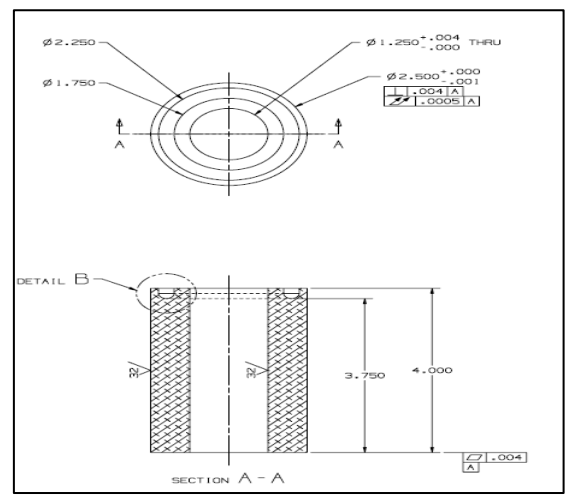

Figure 2: Plug weld design.

| Coupon ID: | MC75 | MC76 | MC77 | MC78 | MC79 | MC80 | AF81 | AF82 | AF83 | AF84 | MC5 | MC6 |
|---|---|---|---|---|---|---|---|---|---|---|---|---|
| Drawing numbers | F10149391 F10149387 F10149388 | F10149391 F10149387 F10149388 | F10149391 F10149387 F10149388 | F10149391 F10149387 F10149388 | F10149391 F10149387 F10149388 | F10149391 F10149387 F10149388 | F10124967 F10124708 F10124959 | F10124967 F10124708 F10124959 | F10124967 F10124708 F10124959 | F10124967 F10124708 F10124959 | F10124967 F10124708 F10124959 | F10124967 F10124708 F10124959 |
| Component/filler rod prep per ED0000862: | Y | Y | Y | Y | Y | Y | Y | Y | Y | Y | Y | Y |
| Electrode size: | 5/32" | 5/32" | 5/32" | 5/32" | 5/32" | 5/32" | 5/32" | 5/32" | 5/32" | 5/32" | 5/32" | 5/32" |
| Electrode type: | 2% Thoriated (CK world wide) | 2% Thoriated (CK world wide) | 2% Thoriated (Weldmark) | 2% Thoriated (Weldmark) | 0.8% Zirconiated (CK world wide) | 0.8% Zirconiated (CK world wide) | 2% Thoriated (Weldmark) | 2% Thoriated (Weldmark) | 2% Thoriated (Weldmark) | 2% Thoriated (Weldmark) | 2% Thoriated (Weldmark) | 2% Thoriated (Weldmark) |
| Truncated tip with 35 deg. Angle? | Y | Y | Y | Y | Y | Y | Y | Y | Y | Y | Y | Y |
| Pre-heat to 230 F? | N | Y | N | Y | N | Y | Y | Y | Y | Y | Y | Y |
| Depth of groove: | 0.25" | 0.25" | 0.325" | 0.325" | 0.25" | 0.25" | -- | -- | -- | -- | -- | -- |
| Filler rod type: | E4043 | E4043 | E4043 | E4043 | E4043 | E4043 | E4043 | E4043 | E4043 | E4043 | E4043 | E4043 |
| Filler rod size: | 1/16" | 1/16" | 1/16" | 1/16" | 1/16" | 1/16" | 1/16" | 1/16" | 1/16" | 1/16" | 1/16" | 1/16" |
| Gas type: | 75 He./25 Ar. (99.9995% HP) | 75 He./25 Ar. (99.9995% HP) | 75 He./25 Ar. (99.9995% HP) | 75 He./25 Ar. (99.9995% HP) | 75 He./25 Ar. (99.9995% HP) | 75 He./25 Ar. (99.9995% HP) | 75 He./25 Ar. (99.9995% HP) | 75 He./25 Ar. (99.9995% HP) | 75 He./25 Ar. (99.9995% HP) | 75 He./25 Ar. (99.9995% HP) | 75 He./25 Ar. (99.9995% HP) | 75 He./25 Ar. (99.9995% HP) |
| Gas flow rate: | 130 SCFH He. | 130 SCFH He. | 130 SCFH He. | 130 SCFH He. | 130 SCFH He. | 130 SCFH He. | 130 SCFH He. | 130 SCFH He. | 130 SCFH He. | 130 SCFH He. | 130 SCFH He. | 130 SCFH He. |
| Root pass: | 1 | 1 | 1 | 1 | 1 | 1 | 1 | 1 | 1 | 1 | 1 | 1 |
| Filler pass: | 2 | 2 | 2 | 2 | 2 | 2 | 3 | 3 | 3 | 3 | 2 | 2 |
| Current and polarity type: | AC with 80% EN | AC with 80% EN | AC with 80% EN | AC with 80% EN | AC with 80% EN | AC with 80% EN | AC with 80% EN | AC with 80% EN | AC with 80% EN | AC with 80% EN | AC with 80% EN | AC with 80% EN |
| Max current setting: | 250 A | 250 A | 250 A | 250 A | 250 A | 250 A | 250 A | 250 A | 250 A | 250 A | NA | NA |
| Frequency setting: | 130 Hz | 130 Hz | 130 Hz | 130 Hz | 130 Hz | 130 Hz | 130 Hz | 130 Hz | 130 Hz | 130 Hz | NA | NA |
| Weld start time: | 9:04 AM | 10:53 AM | 9:21 AM | 11:08 AM | 9:45 AM | 10:20 AM | 9:06 AM | 10:34 AM | 1:30 PM | 8:31 AM | NA | NA |
| Weld stop time: | 9:20 AM | 10:58 AM | 9:27 AM | 11:13 AM | 9:48 AM | 10:27 AM | 9:30 AM | 10:56 AM | 1:52 PM | 8:47 AM | NA | NA |
| Measured temperature prior to root pass: | 68 F | 230 F | 70 F | 202 F | 80 F | 230 F | 230 F | 230 F | 230 F | 250 F | NA | NA |
| Avg. voltage measured root pass: | 20 V | 18 V | 20 V | 21 V | 21.5 V | 21.4 V | 21 V | 21.9 V | 22.2 V | 22.5 V | NA | NA |
| Avg. voltage measured filler pass#1: | 25 V | 20 V | 21 V | 17 V | -- | 20 V | 22 V | 20.4 V | 20.9 V | 20.4 V | NA | NA |
| Avg. voltage measured filler pass#2: | 21 V | 18 V | 20 V | 19 V | 20 V | -- | 20.4 V | 21.4 V | 20.5 V | 20.1 V | NA | NA |
| Avg. voltage measured filler pass#3: | NA | NA | NA | NA | NA | NA | 21 V | 20.7 V | 21.4 V | 20.4 V | NA | NA |
| Avg. current measured root pass: | 248 A | 243 A | 248 A | 250 A | 249 A | 250 A | 230 A | 250 A | 250 A | 250 A | NA | NA |
| Avg. current measured filler pass#1: | 248 A | 235 A | 248 A | 180 A | -- | 230 A | 215 A | 230 A | 230 A | 235 A | NA | NA |
| Avg. current measured filler pass#2: | 243 A | 189 A | 248 A | 200 A | 225 A | 230 A | 214 A | 230 A | 211 A | 215 A | NA | NA |
| Avg. current measured filler pass#3: | NA | NA | NA | NA | NA | NA | 220 A | 200 A | 196 A | 210 A | NA | NA |
| Passed ASME B31.3 Normal Fluid Serv. | Y | Y | Y | Y | Y | Y | Y | Y | Y | Y | Y | Y |

Figure 3: Critical welding parameters recorded during the welding of the coupons.

The above samples are cut at 0.5 in. from the face of the weld and sent for radiography. All the critical welding parameters were documented. These values are highlighted in Fig 3. The thickness of the electrode, type of the filer rod, initial settings on the welding machine, and the shielding gas type and flow rates were kept constant for all the coupons. The type and manufacturer of the electrode, number of filler passes, and the pre-heat values were varied. The coupons are shown in Fig 4 and their radiographs are shown in Fig 5.

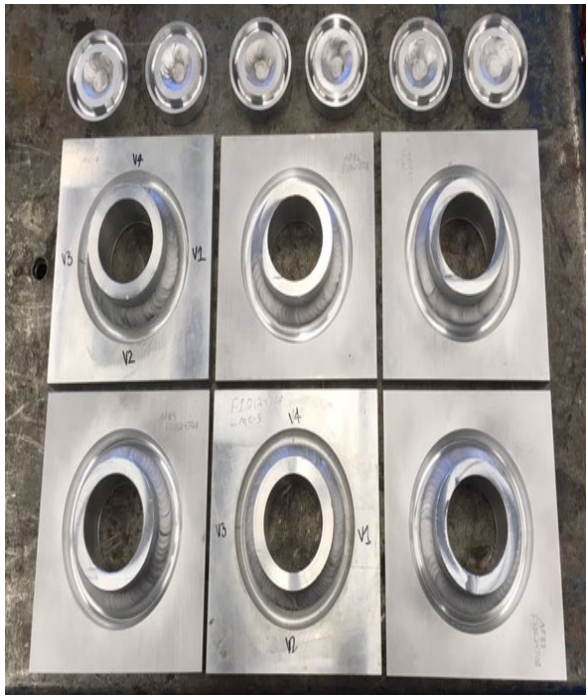

Figure 4: All plug weld and pipe-block weld coupons.

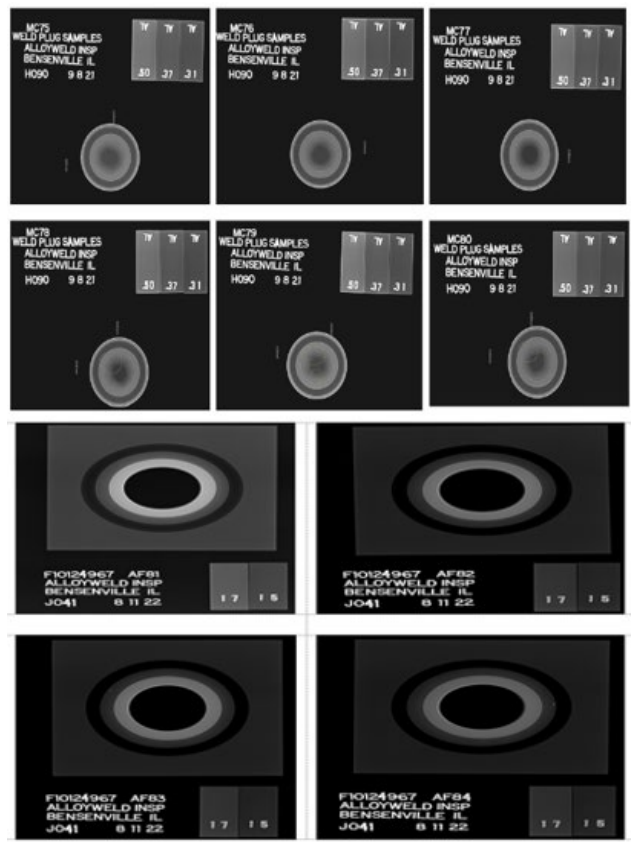

Figure 5: Radiographs of weld coupons shown in Fig 4.

*Effect of Pre-heating and Electrode Size*

Prior to creating the above coupons, few preliminary welding coupons, pipe-block joint, were generated using 1/8 in., 2% thoriated tungsten electrodes. None of the joints were pre-heated. All these samples failed as per NAS 1514 Class I [1] though the WPS was followed meticulously. The coupons had unacceptable amounts of tungsten inclusions in them. A few changes were made to the WPS, namely, the tungsten electrode size was increased to 5/32 in. and one of the weld joints was pre-heated. Both coupons passed per NAS 1514 Class I [1].

To weld the plug weld coupons, they were loaded into a special fixture. The welding process started with loading up the assembly to create coupons MC71 and MC72. MC71 was welded up first and then soon after MC72 was welded. While MC71 was being welded, MC72 warmed up owing to its thermal contact with MC71. In effect, it was artificially pre-heated. The same happened with the MC73 and MC74. The bright spots in the radiographs below are the tungsten inclusions. One can clearly see that the number of tungsten inclusions in the unintentionally pre-heated samples are considerably lower than those in the un-pre-heated ones. This shows that pre-heating the samples prior to welding may have a significant effect on the number of tungsten inclusions in the welds. The comparison of tungsten inclusions in highlighted in Fig 6.

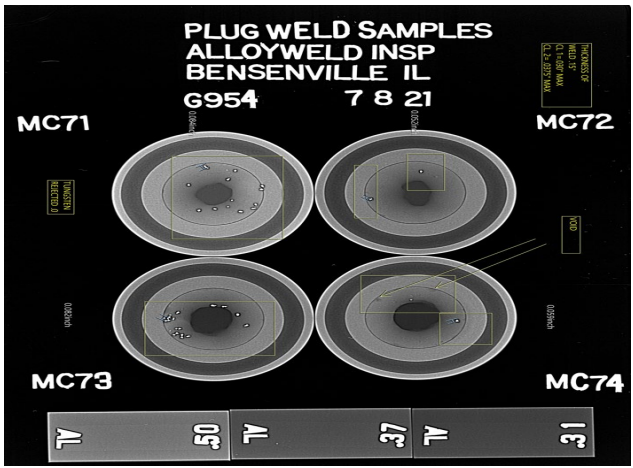

Figure 6: Comparison of tungsten inclusions in pre-heated and cold coupons.

*Weld Defect Comparison with Piping Code*

The weld designs in Fig 1 and Fig 2 are fillet welds, and as per ASME B31.3 [3], they include socket, seal, attachment welds for slip-on flanges, branch reinforcement, and supports. The piping and plug connections are part of the fluid circuit which is connected to the Absorber RAW cooling system. The design of the cooling system conforms to ASME B31.3 Normal Fluid Service [3]. Thus, it seemed appropriate to subject the coupons to the piping code's weld qualification criteria. Thus, all the coupons were sent for re-evaluation. All the samples passed as per ASME B31.3 Normal Fluid Service [2]. These reports are highlighted in Fig 7.

## CONCLUSION

To produce production quality welds, a proven base WPS must be followed. Larger diameter tungsten electrodes, 5/32 in., with pre-heated assemblies result in better quality welds. This could be because the thinner electrodes breakdown easily and get lodged into the weld pool during the welding process. This breakdown is further enhanced by the large temperature differential between the un-pre-heated sample and the hot electrode. Both pre-heated and un-preheated coupons passed the piping code examination criteria. It was observed that the type and manufacturer of the tungsten electrode had little to no effect on the quality of the weld produced.

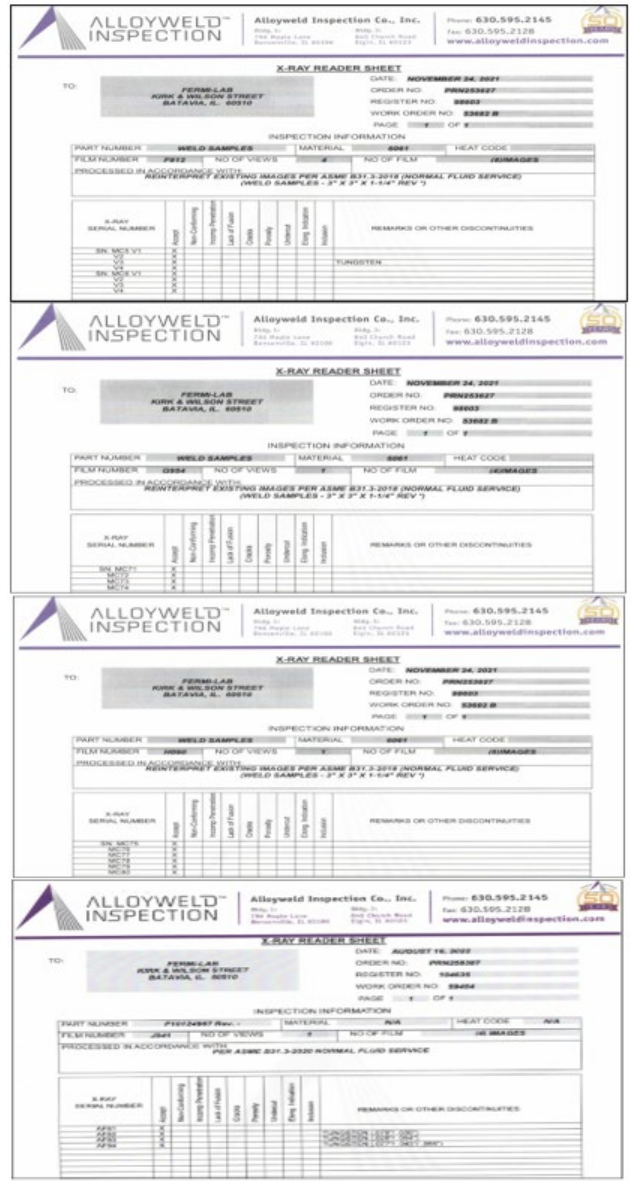

Figure 7: Weld coupon radiography reports.

# REFERENCES

[1] American Society of Mechanical Engineers B31.3 "Process piping an American national standard", 2022, American Society of Mechanical Engineers, New York, NY 10016-5990.

[2] AIA/NAS-NAS 1514, "Radiographic standard for classification of fusion weld discontinuities", 2015, AIA/NAS Aerospace Industries Association, Arlington, VA.

[3] B. D. Hartsell *et al*., "LBNF hadron absorber: Mechanical design and analysis for 2.4 MW operation", in *Proc. IPAC'15*, Richmond, VA, USA, May 2015, pp. 3318-3320. doi:10.18429/JACoW - IPAC2015 - WEPTY025

[4] A. Deshpande *et al.*, "LBNF hadron absorber: Updated mechanical design and analysis for 2.4 MW operation", in *Proc. IPAC'19*, Melbourne, Australia, May 2019, pp. 4078-4080. doi:10.18429/JACoW - IPAC2019 - THPRB108